\documentclass[letterpaper]{article} 
\usepackage{aaai24}  
\usepackage{times}  
\usepackage{helvet}  
\usepackage{courier}  
\usepackage[hyphens]{url}  
\usepackage{graphicx} 
\usepackage{natbib}  
\usepackage{caption} 
\usepackage{algorithm}
\usepackage{algorithmic}
\usepackage{url}
\usepackage{xcolor} 

\usepackage{todonotes}

\usepackage{newfloat}
\usepackage{listings}
\DeclareCaptionStyle{ruled}{labelfont=normalfont,labelsep=colon,strut=off} 
\floatstyle{ruled}
\newfloat{listing}{tb}{lst}{}
\floatname{listing}{Listing}
\usepackage{booktabs}
\usepackage{array}

\newcolumntype{M}[1]{>{\centering\arraybackslash}m{#1}}

\title{Teenagers and Artificial Intelligence: Bootcamp Experience and Lessons Learned} 
\author{
    Uzay Macar,
    Blake Castleman,
    Noah Mauchly,
    Michael Jiang,
    Asma Aouissi,
    Salma Aouissi,
    Xena Maayah,
    Kaan Erdem,
    Rohith Ravindranath,
    Andrea Clark-Sevilla,
    Ansaf Salleb-Aouissi
}
\affiliations{
    Aiphabet Inc. \\
    contact@aiphabet.org
}

\begin{document}

\maketitle

\begin{abstract}

Artificial intelligence (AI) stands out as a game-changer in today's technology landscape. However, the integration of AI education in classroom curricula currently lags behind, leaving teenagers inadequately prepared for an imminent AI-driven future. 
In this pilot study, we designed a three-day bootcamp offered in the summer of 2023 to a cohort of 60 high school students. The curriculum was delivered in person through animated video content, easy-to-follow slides, interactive playgrounds, and quizzes. These were packaged in the early version of an online learning platform we are developing.
Results from the post-bootcamp survey conveyed a 91.4\% overall satisfaction. Despite the short bootcamp duration, 88.5\% and 71.4\% of teenagers responded that they had an improved understanding of AI concepts and programming, respectively. Overall, we found that employing diverse modalities effectively engaged students, and building foundational modules proved beneficial for introducing more complex topics. Furthermore, using Google Colab notebooks for coding assignments proved challenging to most students. Students' activity on the platform and their answers to quizzes showed proficient engagement and a grasp of the material.
Our results strongly highlight the need for compelling and accessible AI education methods for the next generation and the potential for informal learning to fill the gap of providing early AI education to teenagers.

\medskip
\textbf{Keywords}: AI education, AI curriculum, AI bootcamp, online learning platform, informal learning, teenagers and AI. 

\end{abstract}

\begin{figure*}[ht]
\centering
\fbox{\includegraphics[scale=0.43]{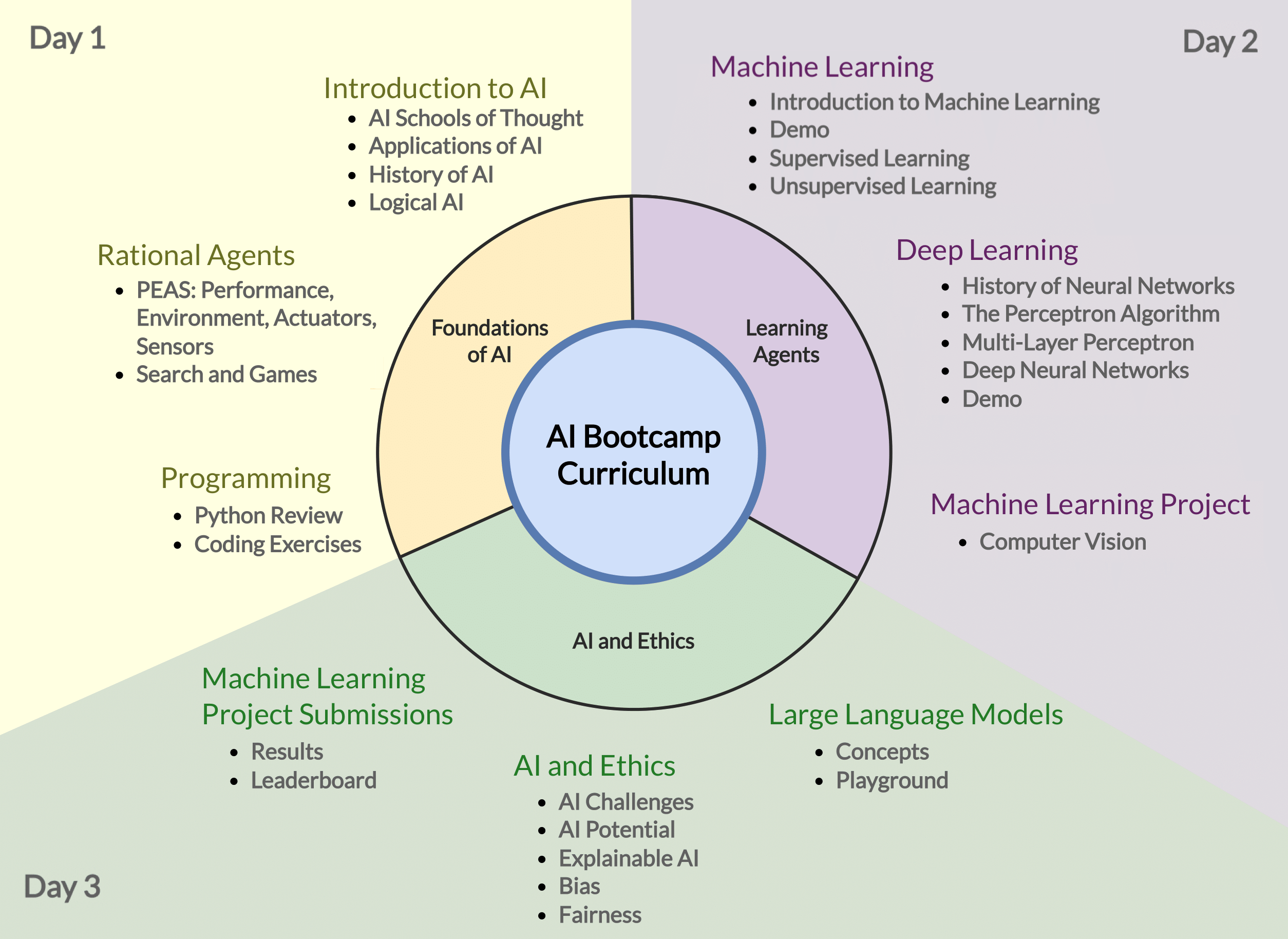}}
\caption{The curriculum presented to students in the three-day AI bootcamp}
\label{curriculum:overview}
\end{figure*}

\section{Introduction}

We are witnessing an incredible leap in artificial intelligence (AI) research and applications, affecting every aspect of our lives. As AI continues to shape and revolutionize numerous domains, addressing the existing knowledge gap is crucial, particularly among teenagers \cite{imagineEthicalAI}. Youth’s perceptions of AI are often encompassed by science fiction and popular culture \cite{Greenwald_Leitner_Wang_2021}, leading to common misconceptions. It is critical to create powerful learning opportunities to empower teenagers with the necessary skills and ethical awareness to navigate the AI-driven world responsibly, harness its potential, and address the challenges that arise from its adoption.

Growing motivation and evidence exist for an early AI education installation adapted to young learners before college. There is also a strong interest from different stakeholders, including academia, industry, and governments. Different institutions have called for designing a K-12 AI education agenda on how to create curriculum that is engaging from both the technical and ethical perspectives \cite{miao2022k,Zhang_Lee_Ali_DiPaola_Cheng_Breazeal_2022}, and prepare an AI-ready workforce \cite{aiforce}.

There is also growing literature in education research and fantastic efforts from the AI in education community (\citeauthor{Schaper2023} \citeyear{Schaper2023};  \citeauthor{lane2023} \citeyear{lane2023};  \citeauthor{Bellas2022AICF} \citeyear{Bellas2022AICF}; \citeauthor{Hasse_Cortesi_Lombana_Bermudez_Gasser_2019} \citeyear{Hasse_Cortesi_Lombana_Bermudez_Gasser_2019}; \citeauthor{Zhang_Lee_Ali_DiPaola_Cheng_Breazeal_2022} \citeyear{Zhang_Lee_Ali_DiPaola_Cheng_Breazeal_2022}; \citeauthor{RIZVI2023100145} \citeyear{RIZVI2023100145}).
Despite rich contributions so far, there is a need for more research to fit AI into formal education in schools. Indeed, there are unique challenges to overcome. These challenges, to name a few, include:
\begin{itemize}
\item Training teachers for AI education in the classroom.
\item Addressing disparities in schools' computing resources.
\item Homogenizing CS education across schools.
\item Fitting AI learning into packed school schedules.
\item Designing AI curriculum that encompasses an appropriate depth level for diverse age ranges. 
\item Figuring out entry points to AI across the K-12 spectrum.  
\item Researching and standardizing AI content and evaluation framework for K-12.
\end{itemize}

Prior literature from the AI4K12 team \cite{Touretzky2023,touretzky2019special,touretzky2021artificial} provide a useful framework, called the ``five big ideas.'' The authors propose guidelines on how to structure AI curricula around perception, representation and reasoning, learning, natural interaction, and societal impact for different K-12 age groups. While this framework is adopted, AI's full integration into school curriculum will take time.

Perhaps, as pointed out by \cite{lane2023}, a realistic approach is to strengthen the K-12 curriculum to provide the AI prerequisites in school thus preparing students for college. This would include a more robust mathematical background, ideally including discrete mathematics, proofs and critical thinking, calculus, linear algebra, probability, and statistics. It also means enforcing stronger computer science (CS) preparedness \cite{kunda2021triplet}.

The aforementioned evidence highlights the need to provide opportunities for informal AI education outside the traditional classroom. Our methodology aligns with this burgeoning research domain, which presents copious unresolved challenges. These include deciding the best cutoff points to group K-12 learners into sensible ranges, crafting informal AI curriculum tailored for each distinct age group, conducting rigorous educational research with empirical evidence and assessment of learning outcomes, and finding the entry points to make AI education suitable for different K-12 age groups in an informal setting.

Our organization's contribution lies in the sphere of informal AI learning and contributes with an exhaustive curriculum encompassing the various disciplines of AI (Figure \ref{curriculum:overview}). With diverse modalities including but not limited to video lectures, interactive playgrounds, applied programming labs, and quizzes, we seek to create curricula inclusive of all learning styles and ages. We conjecture that AI informal education is the most feasible approach to make AI education more accessible to young learners in the near future while giving time for formal learning to mature and set the proper scaffolding from all perspectives to make AI education a fundamental subject in schools. 

In this paper, we target specifically teenage learners to empower them with informal on-demand AI education. We developed a pilot study in the form of a three-day bootcamp offered in the summer of 2023 to a cohort of 60 high school students. The bootcamp curriculum was delivered in person through animated video content, user-friendly slides, interactive playgrounds, and quizzes. The bootcamp material is packaged from the early version of our organization's online learning platform actively being developed.

We present the various lessons learned from this experience. Overall, we found that employing diverse modalities effectively engaged students, and building foundational modules proved beneficial for introducing more complex topics. Students' answers to quizzes showed proficient engagement and a grasp of the material. Incorporating live chat boosted participation. Finally, using Google Colab notebooks for coding assignments proved challenging to most students. This finding prompts the need for a suitable coding interface to teach AI and programming to teenagers.

A high-level overview of our pilot study is provided in Table \ref{bootcamp:overview}. This study forms the first milestone of a larger non-profit project, with a mission to bring high-quality informal AI education to teenagers outside the classroom.

\begin{table}[h]
\def\arraystretch{1.43}
\centering
\begin{tabular}{|m{1.75cm}|m{5.65cm}|}
 \hline
 \textbf{What} & Comprehensive AI bootcamp\\
 \hline
 \textbf{Audience} & 60 high school students\\
 \hline
 \textbf{Instructors} & A university CS professor and a domain-expert teaching assistant \\
 \hline
 \textbf{Medium} & Hybrid: in-person and online platform\\
 \hline
 \textbf{Duration} & 3 days of 4 hour sessions \\
 \hline
  \textbf{Metrics} & Survey results, completion rates, and quiz grades \\
 \hline
\end{tabular}
\caption{A high-level overview of our study}
\label{bootcamp:overview}
\end{table}


\section{Cohort}
\label{cohort}


The cohort consisted of $60$ high school students aged 14-19 years old. Students came from several different high schools in (redacted information) who signed up to the bootcamp prior through an online form. We reached out to the relevant population through different marketing channels roughly a month prior, and found email newsletters and Facebook groups to be the most effective channels. The marketing material consisted of a high-level overview of the bootcamp and the curriculum to be covered.

\begin{figure}[h]
\centering
\includegraphics[scale=0.65]{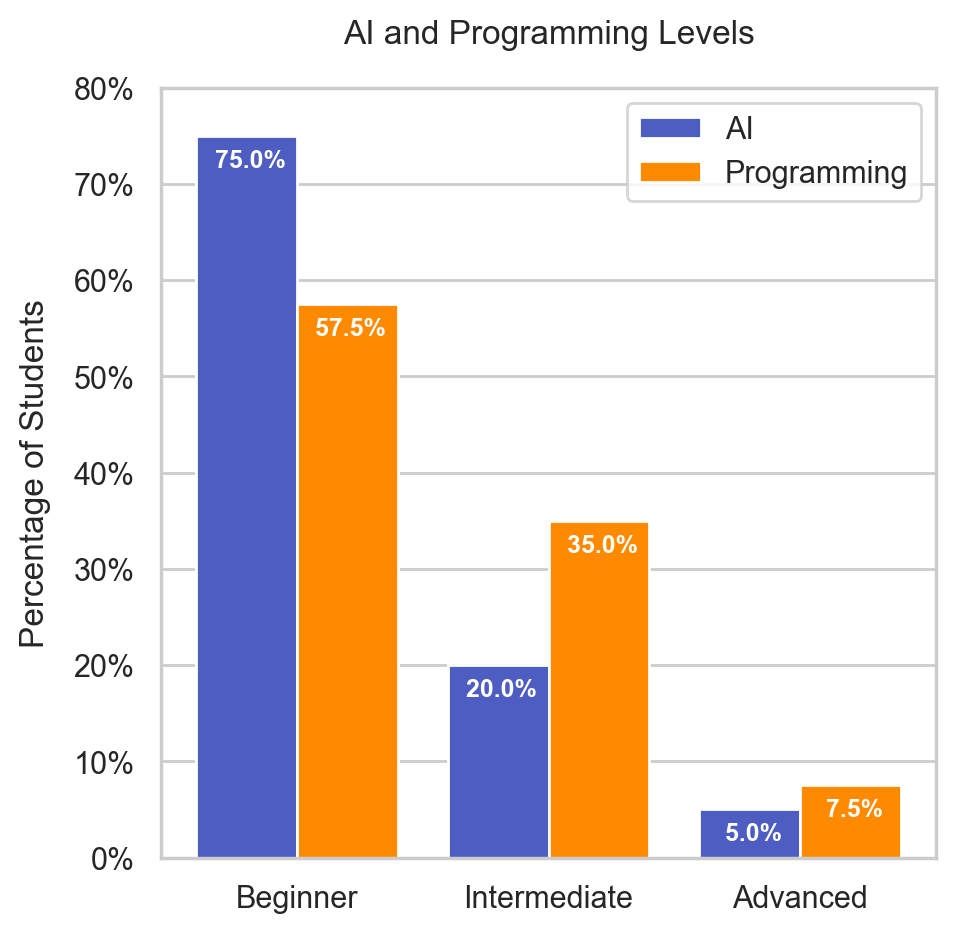}
\caption{Students' prior levels in AI and programming}
\label{cohort:level}
\end{figure}

To get to know the student profiles, we conducted a pre-questionnaire at the beginning of the bootcamp. We found that among the students who signed up the bootcamp, mathematics and physics were the two most popular courses in the high school curriculum. When comparing AI and programming proficiency levels (Figure \ref{cohort:level}), we observe that more students are beginners in AI whereas programming proficiency shows a higher percentage of students in the intermediate and advanced levels.

\section{Curriculum}
\label{curriculum}

We designed a comprehensive AI curriculum that balances the foundational concepts with an understanding of the state-of-the-art, organized around three concepts (Figure \ref{curriculum:overview}): 
\begin{itemize}
  \item Day 1: Foundations of AI
  \item Day 2: Learning Agents
  \item Day 3: AI and Ethics
\end{itemize}

The first day kicked off with an introduction to AI, featuring applications of AI, history, and logical agents. The history of AI was emphasized to help situate the students on the current state of intelligent systems. Rational agents, the most common approach to building AI agents today \cite{RN2020}, were covered next. This part was comprised of an overview of search agents: simple search, adversarial search, and constraint satisfaction problems (CSPs). We illustrated search algorithms through well-known games students are already familiar with, such as chess, Sudoku, and checkers. We ended the first day with introductory Python exercises, where students formed groups to solve the exercises with hands-on assistance from the instructors.

Learning agents were introduced on day two. Starting with the Perceptron algorithm helped set the ground for supervised learning. We leveraged students' high school mathematics background to use composition functions in building multi-layer perceptrons (MLP) and generalized it to neural networks and deep learning. Live demonstrations using Google TensorFlow\footnote{\url{https://playground.tensorflow.org/}} and Teachable Machine\footnote{\url{https://teachablemachine.withgoogle.com/}} reinforced the concepts and provided a playground for students to experiment with image classification. We ended day two by introducing a computer vision project to build a convolutional neural network for the FashionMNIST \cite{xiao2017fashionmnist}, a dataset of $28 \times 28$ grayscale images of $10$ fashion categories. Students used a Google Colab notebook with skeleton code to train, test, and evaluate their models.

At the beginning of day three, we reviewed with students their projects and discussed different ways (e.g. data augmentations, network complexity, hyperparamater tuning) to boost the test performance of their classifiers. We then presented a hands-on coding module where students learned how to use the OpenAI API in Python to programmatically interact with ChatGPT and experiment with prompt engineering. This part ended with students building custom chatbots for a restaurant to assist customers with ordering. We finished the bootcamp with an important component, AI and Ethics. We covered vulnerabilities and common failure modes of machine learning systems, fairness and bias, and how recent research strives for explainable and safe AI. Finally, we held a discussion session on the challenges and potential of AI. All participating students received a certificate at the end of the bootcamp.

\section{Delivery}
\label{delivery}

The curriculum was delivered in a hybrid format, combining an online learning platform for students to follow and engage with the material on their personal devices and an in-person component with instructors presenting the material with a slideshow. The platform was accessible both in mobile and web. There were two instructors: a university CS professor and a domain-expert teaching assistant.

The online learning platform was developed in-house, using React and JavaScript for the front-end and Amazon Web Services for the cloud back-end. Anonymous data on student engagement and activity was collected and analyzed for research purposes. The platform featured a comments section and a one-to-one live chat modal through the Intercom software \cite{intercom}, which we found to be particularly engaging and effective in addressing student questions and technical challenges asynchronously.

\begin{figure}[h]
\centering
\fbox{\includegraphics[scale=0.14]{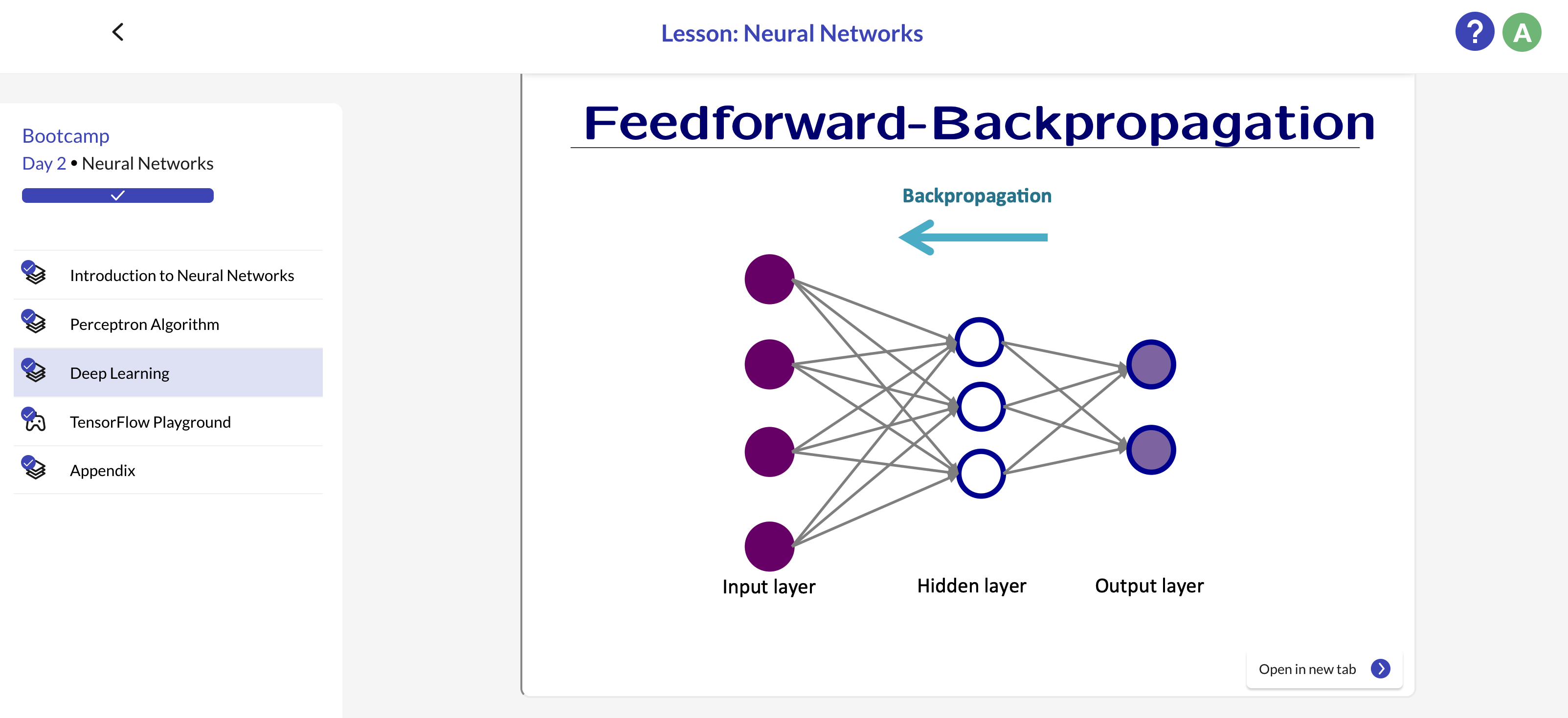} }
\caption{Platform view with a slide on neural networks}
\label{curriculum:slides}
\end{figure}

\begin{figure}[h]
\centering
\fbox{\includegraphics[scale=0.155]{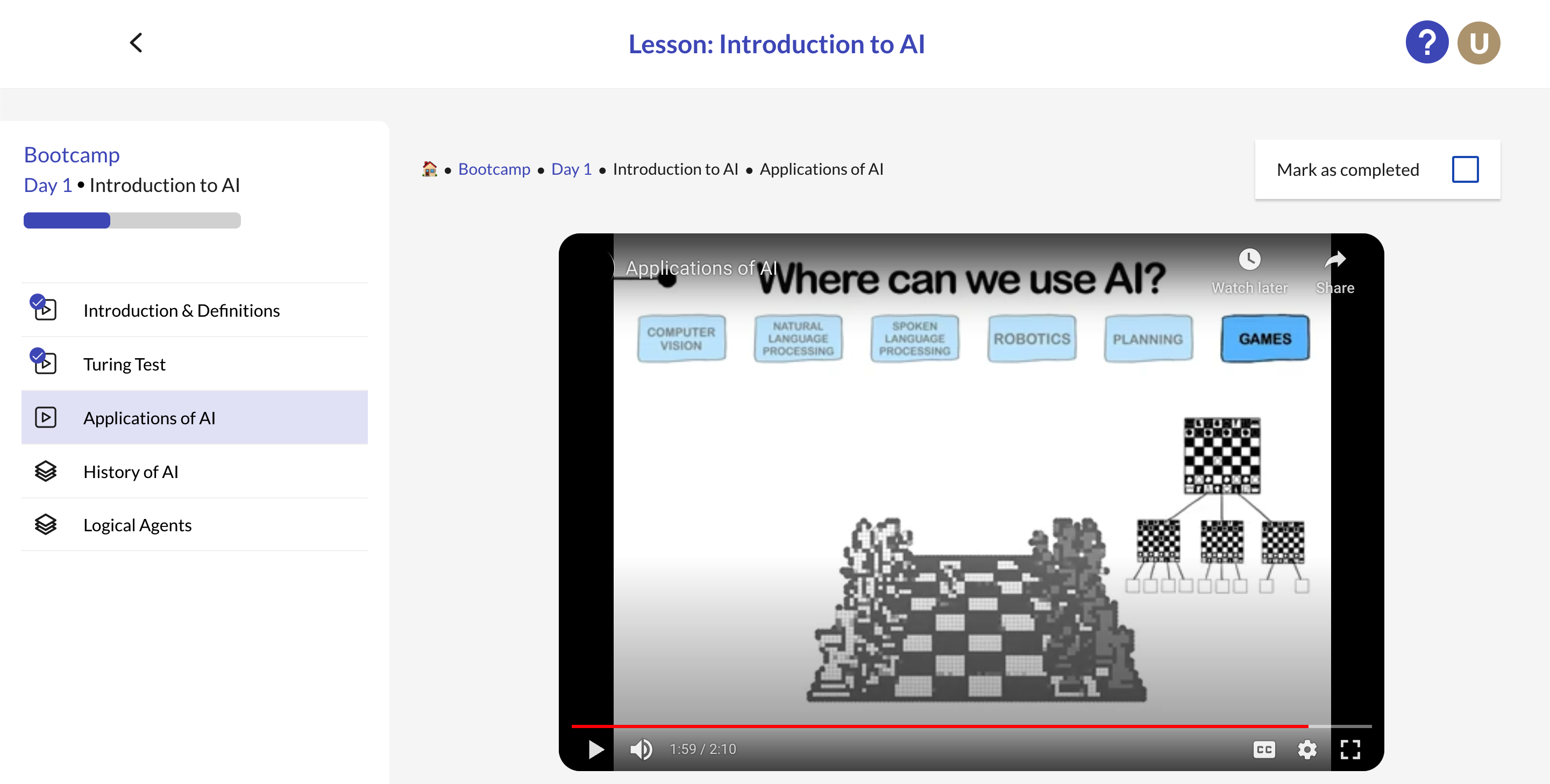} }
\caption{Platform view with a video on AI applications}
\label{curriculum:video}
\end{figure}

\begin{figure}[h]
\centering
\fbox{\includegraphics[scale=0.11]{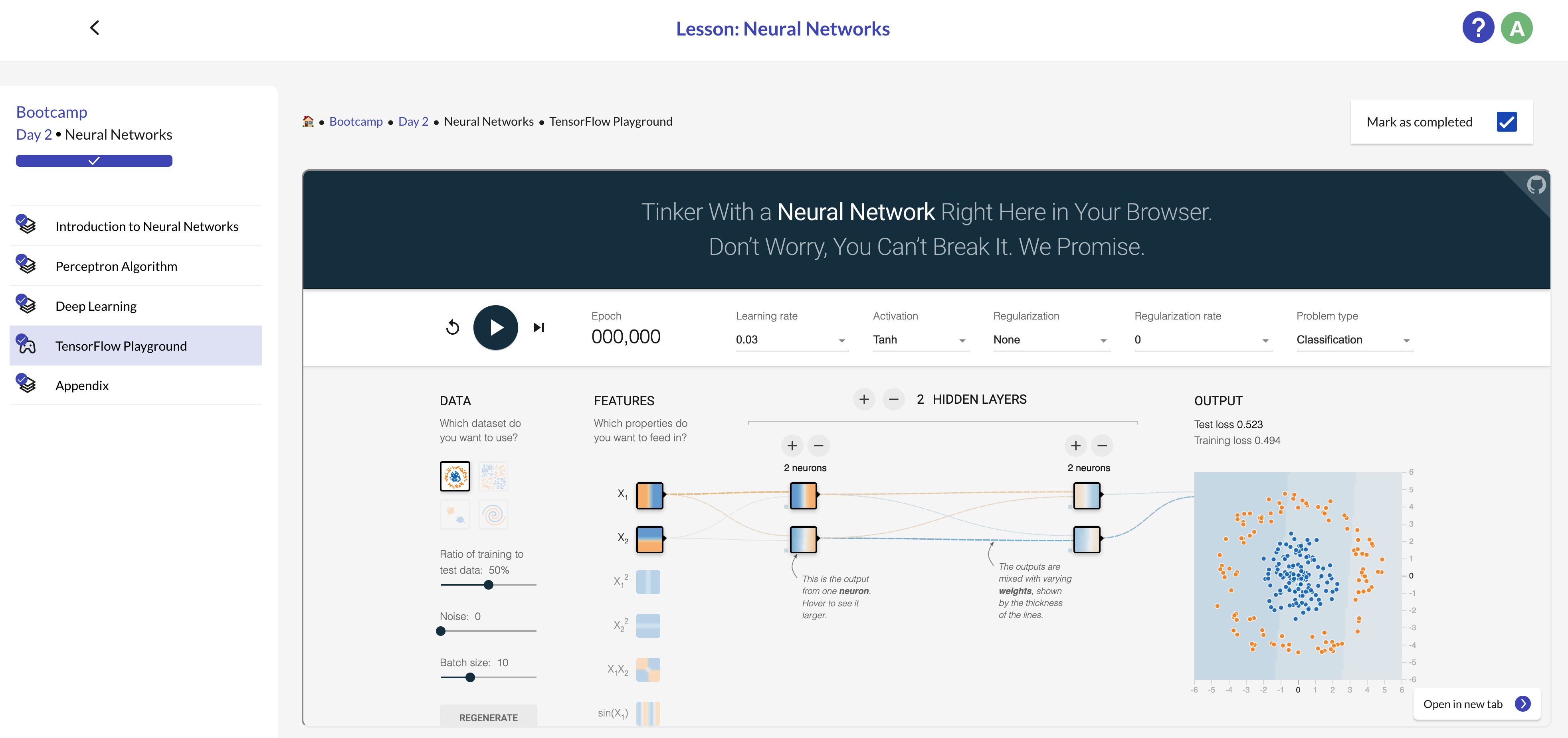} }
\caption{Platform view with a Google TensorFlow demo}
\label{curriculum:demos}
\end{figure}

\begin{figure}[ht]
\centering
\fbox{\includegraphics[scale=0.31]{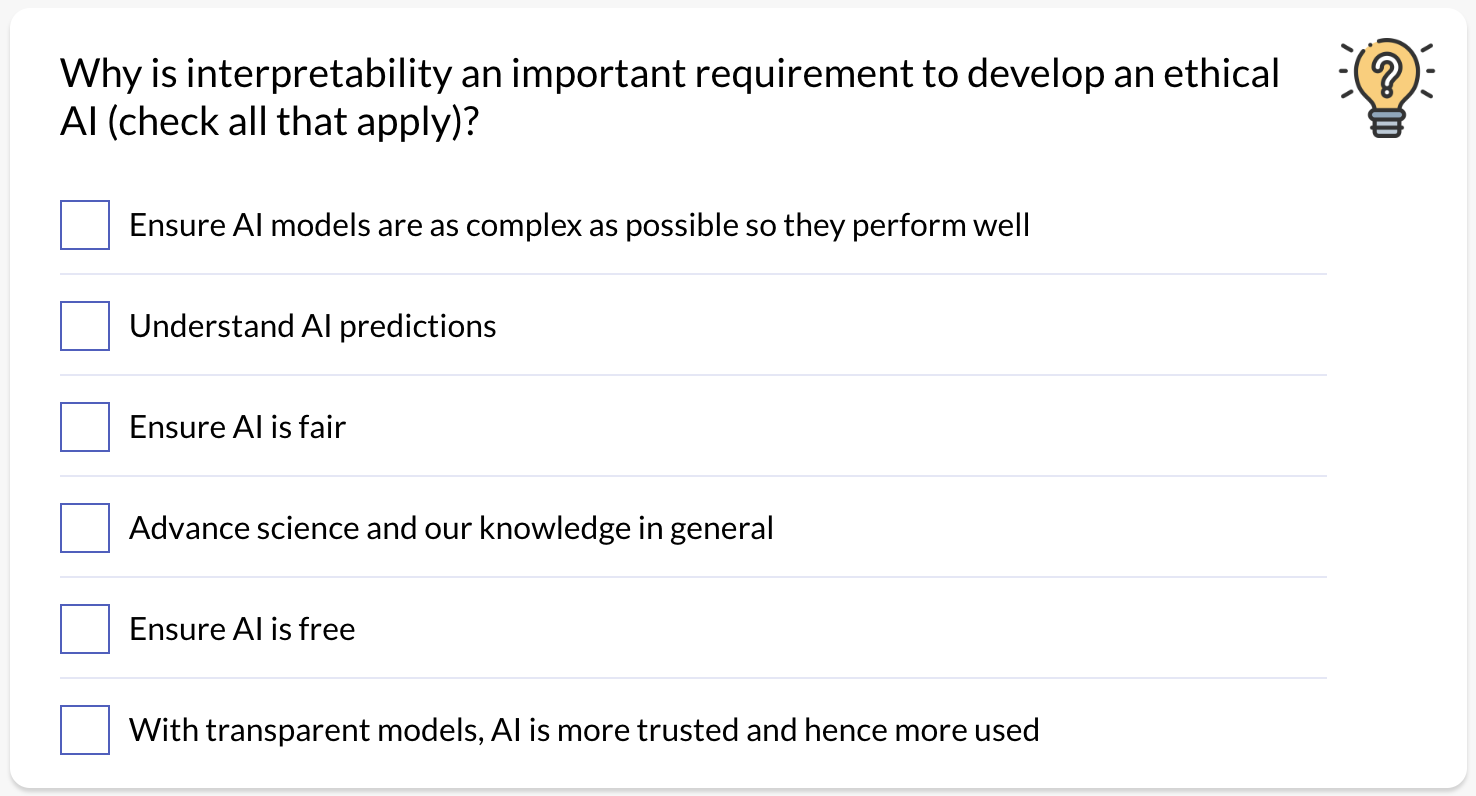}}

\medskip

\fbox{\includegraphics[scale=0.33]{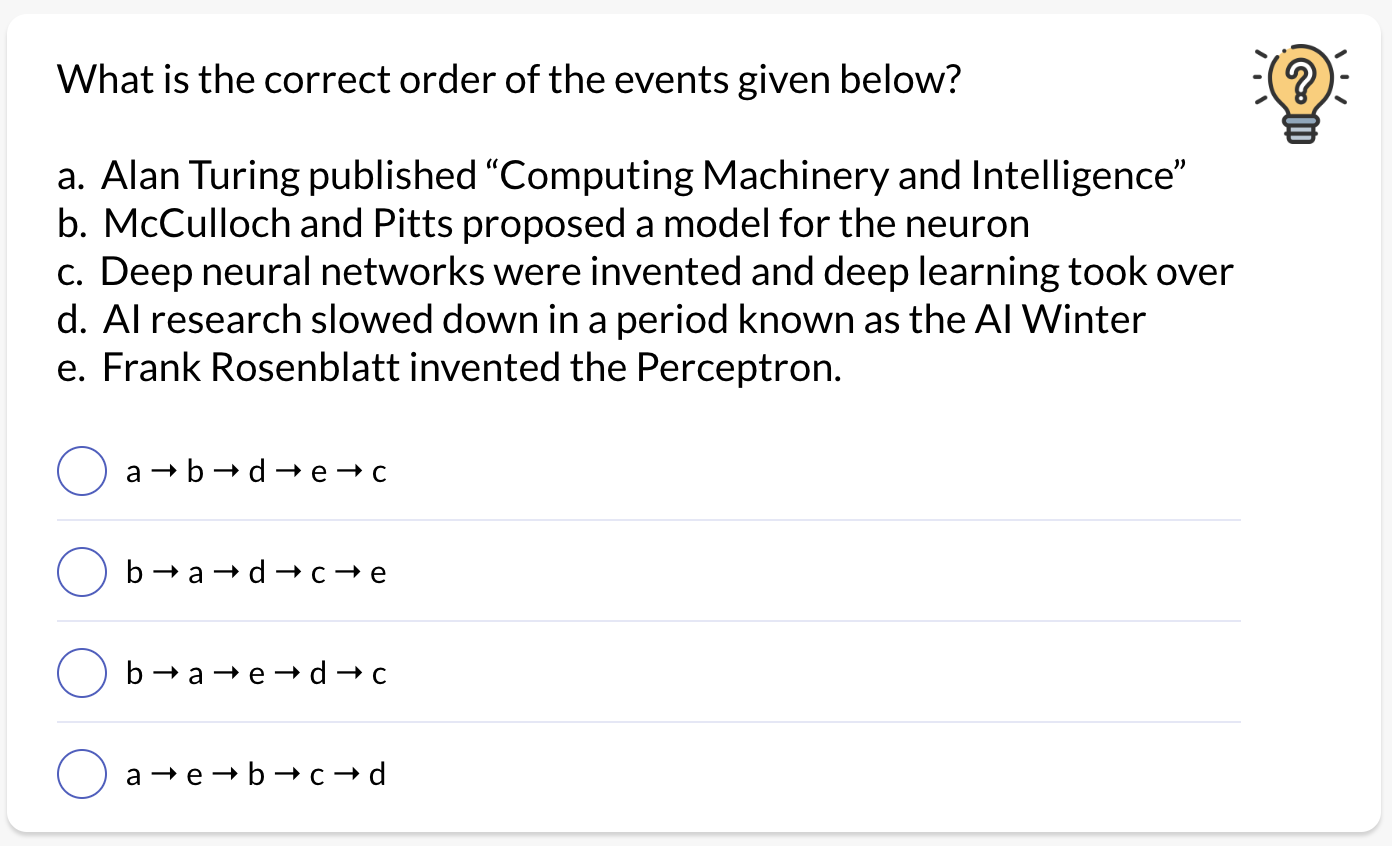}}
\caption{Examples of quiz questions: (Top) AI and Ethics and (Bottom) History of AI}
\label{curriculum:quiz}
\end{figure}

The material was conveyed through a combination of different mediums: slides (e.g. Figure \ref{curriculum:slides}), animated videos (e.g. Figure \ref{curriculum:video}), embedded demos (e.g. Figure \ref{curriculum:demos}), Google Colab notebooks\footnote{\url{https://colab.research.google.com/}}, and quiz questions (e.g. Appendix Figure \ref{curriculum:quiz}). Slides, questions, and scripts for the videos were curated by university professors with assistance from a team of CS undergrads and high school students. Videos, animations, and illustrations were designed by our animation team.

\section{Results}
\label{results}

\begin{figure}[ht]
\centering
\includegraphics[scale=0.6]{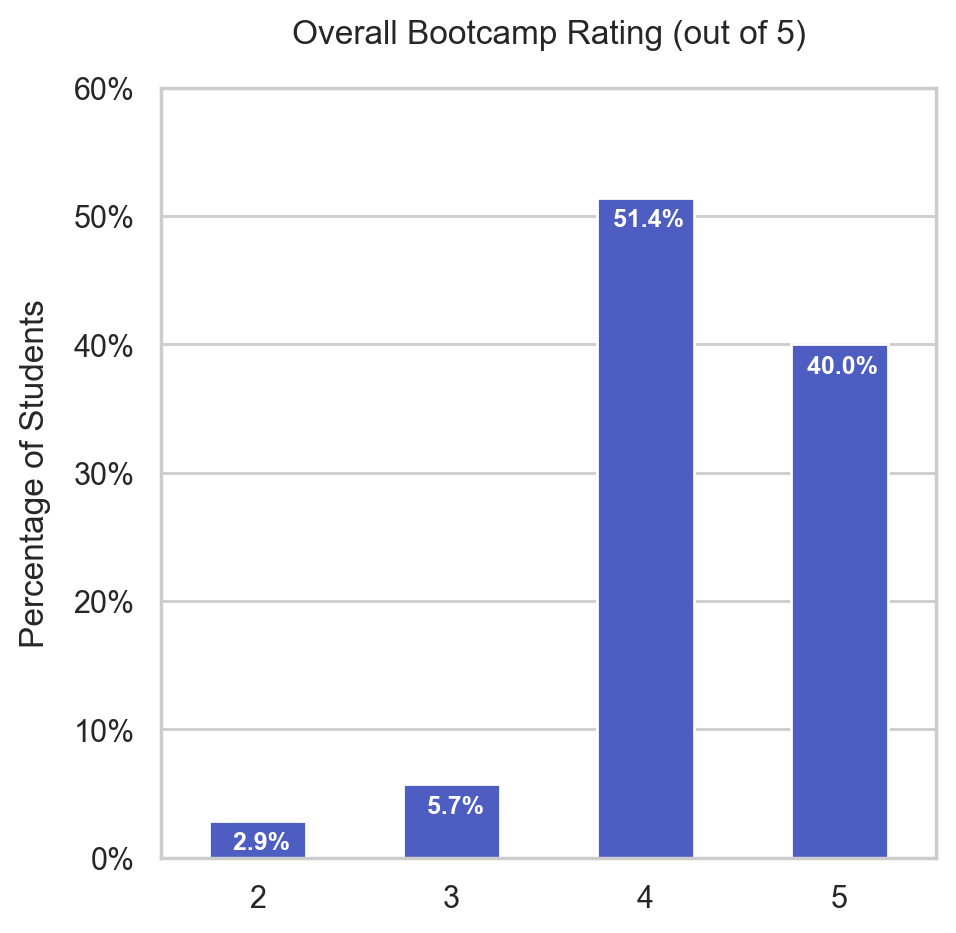} 
\caption{Students' overall rating of the bootcamp} 
\label{rating}
\end{figure}

To assess the effectiveness of our approach, we conducted a post-survey to which $35$ out of $60$ students in the cohort responded. Results from the survey conveyed that $91.4\%$ of students rated the overall bootcamp quality as high ($4$ out of $5$) to very high ($5$ out of $5$) as shown in Figure \ref{rating}. Despite the short bootcamp duration, $88.5\%$ and $71.4\%$ of students responded that they had an improved understanding of AI concepts and programming, respectively.

\begin{figure*}[h]
\centering
\includegraphics[scale=0.6]{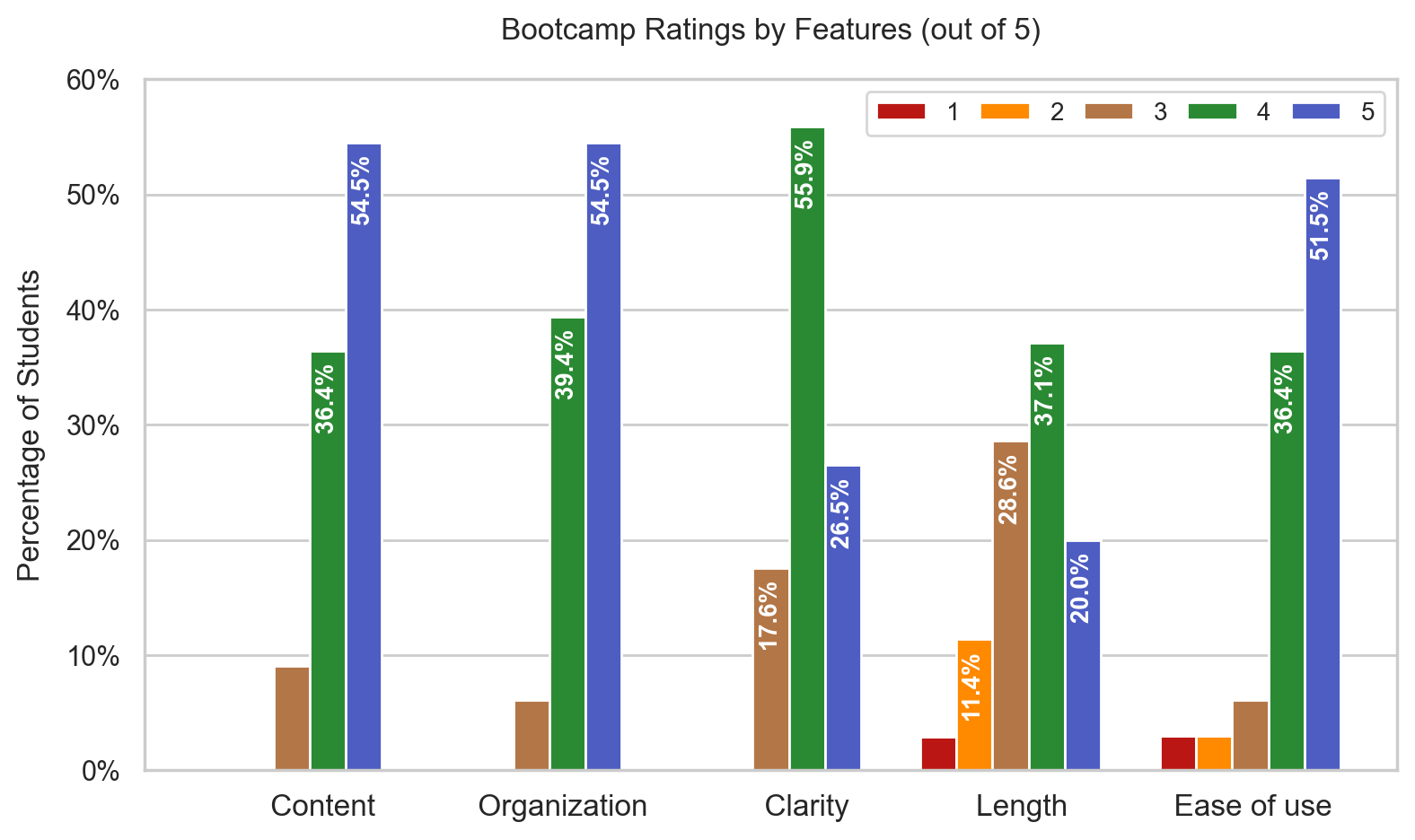}
\caption{Student ratings of the bootcamp by features; percentages smaller than $10\%$ are not labeled} 
\label{rating_category}
\end{figure*}

While the curriculum content, the organization of the bootcamp, and the ease of use of the platform received high ratings, the length of the bootcamp (four hours per day for three days) posed challenges to our students as shown in Figure \ref{rating_category}. Qualitative feedback supports this, with a few students suggesting to increase the number of days and decrease the number of hours per day. Clarity was rated relatively lower as well which we suspect is linked to the delivery of the programming modules with Google Colab notebooks, again supported by qualitative feedback. More observations on this are described in the Discussion section.

\begin{figure*}[ht]
\centering
\includegraphics[scale=0.4]{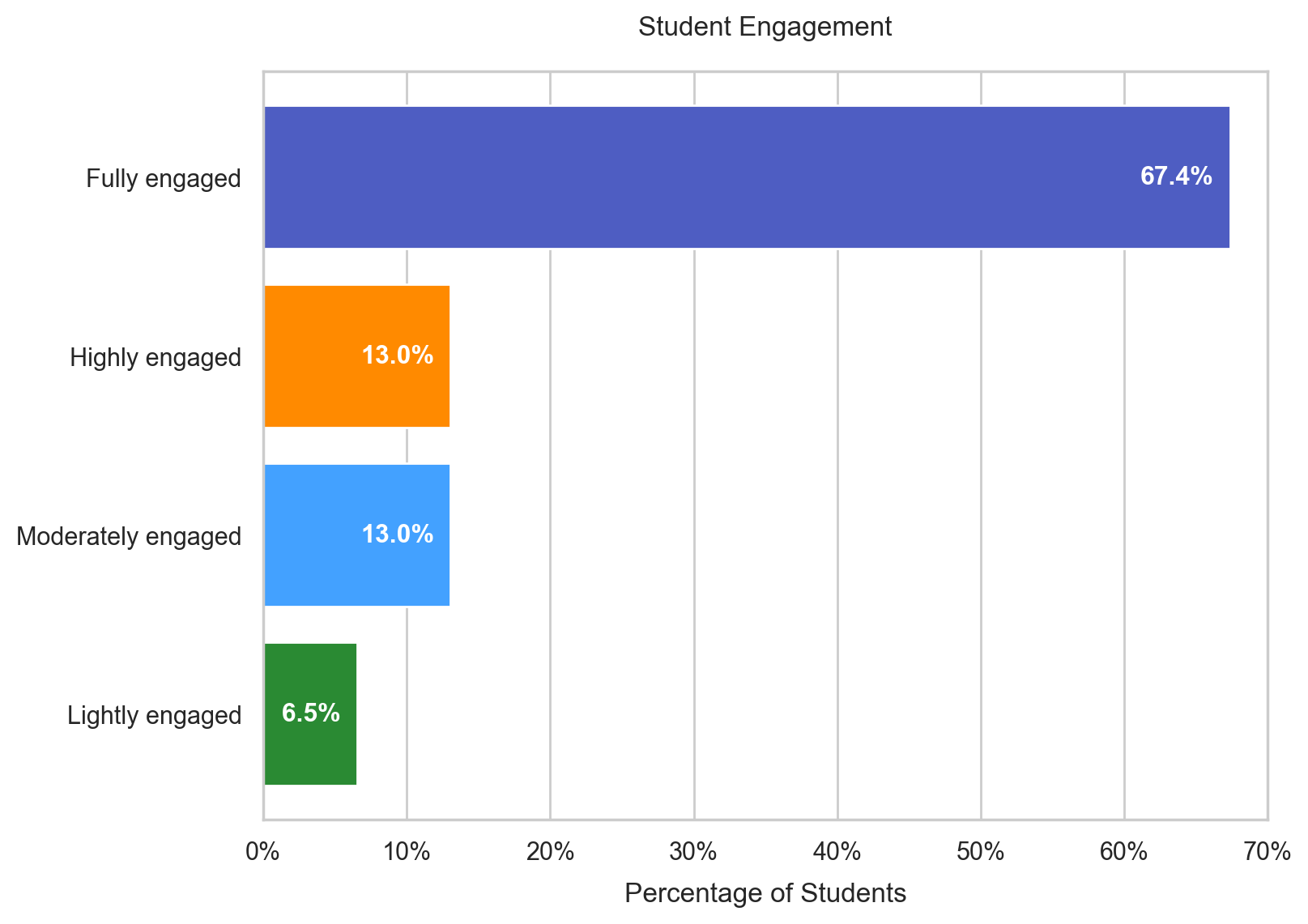} 
\hspace{24pt}
\includegraphics[scale=0.4]{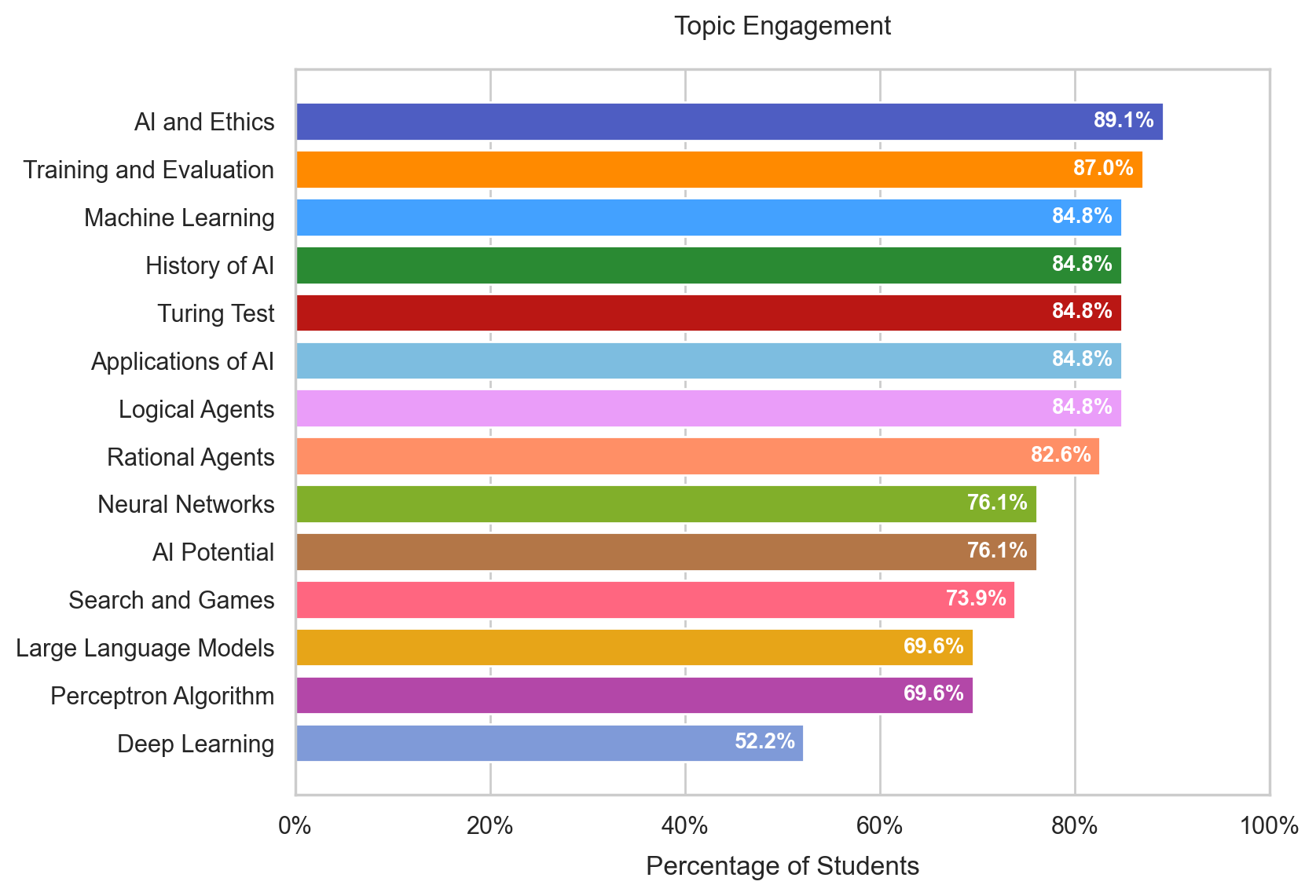}
\caption{(Left) Students' overall engagement and (Right) engagement per topic}
\label{engagement}
\end{figure*}

On the other hand, the bootcamp demonstrated high levels of student engagement based on our in-person observations and proxy data. The students were attentive throughout the bootcamp, actively following the material and asking thoughtful questions. In order to quantitatively measure engagement, we placed a button to mark each topic as completed in the platform, visible at the top right in Figures \ref{curriculum:video} and \ref{curriculum:demos}. This assisted in incorporating indirect gamification techniques into the learning experience. The limitations of this approach are discussed in the Discussion section. Using how many topics students have marked as completed, i.e., the completion rate, as a proxy for engagement, we found that $80.4\%$ of students were highly or fully engaged. For categorization, we picked the following cut-off points by observing the distribution: between $0\%$ and $25\%$ for lightly engaged, $25\%$ and $50\%$ for moderately engaged, $50\%$ to $75\%$ for highly engaged, and $75\%$ to $100\%$ for fully engaged. Looking into engagement per topic, we found that students engaged more with AI and Ethics and foundational topics and less on topics like deep learning and large language models. Details can be found in Figure \ref{engagement}.

\begin{figure}[h]
\centering
\includegraphics[scale=0.65]{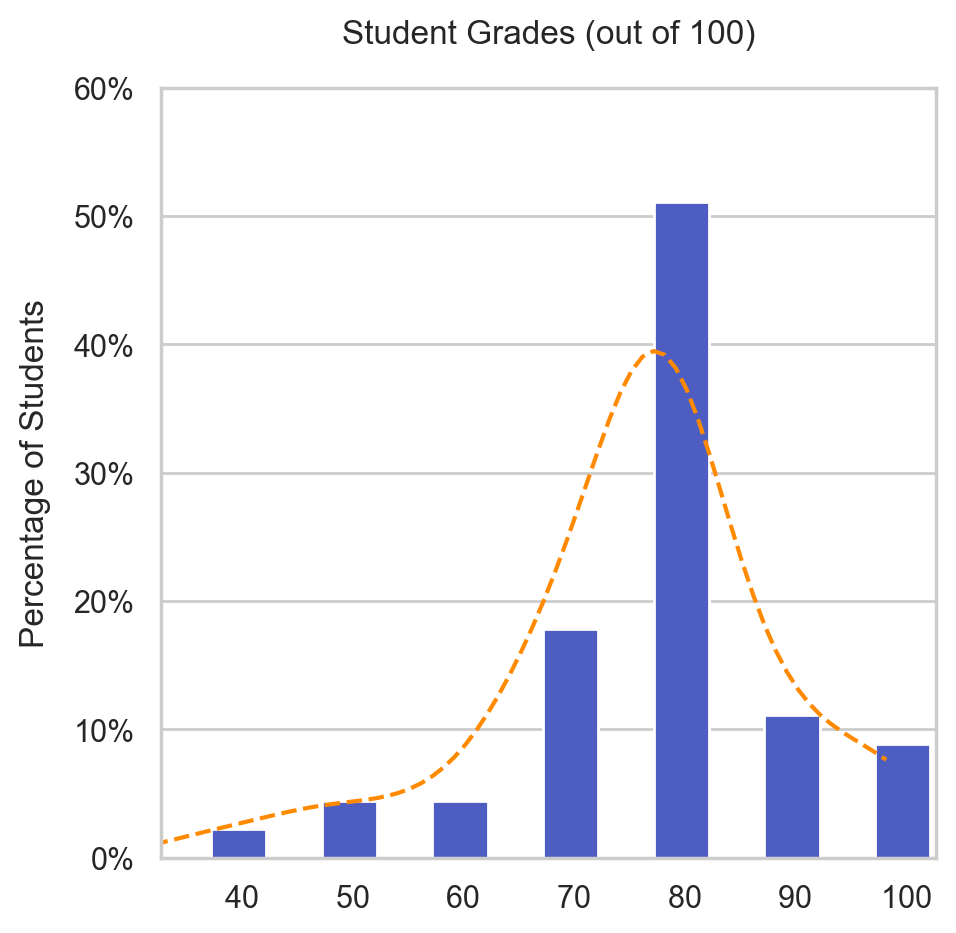}
\caption{Students' quiz grades}
\label{grades}
\end{figure}

Furthermore, students' accurate answers to quiz questions demonstrated a strong grasp of the material. The platform allowed students to submit up to three attempts per question and counted an answer as correct if it was included in these attempts. An overall grade metric for assessment purposes was deducted by dividing the number of questions they answered correctly by the total number of questions. The cohort scored a mean grade of $78.0 \pm 12.4$ out of $100$. Figure \ref{grades} illustrates the grade distribution. 

The quiz questions were distributed amongst diverse topics to maximize the scope of the assessment. Students received the highest grades on foundational topics like rational agents and Turing test. However, students found questions on AI and Ethics and the History of AI challenging. 

\section{Discussion}
\label{discussion}

Given the prevalence of AI in today's world, our research efforts were fueled by a mission to bring high-quality AI education to teenagers outside of the classroom. Our pilot study has identified a significant gap and latent demand in CS and AI education for teenagers, indicating a pressing need for broader education and outreach efforts. 

Our study demonstrated that combining different teaching modalities (e.g. slides, videos, playground, quizzes) proved effective in maintaining student engagement through interactivity and gamification. The bootcamp's hybrid format, combining online and in-person components, provided multiple simultaneous ways for students to interact with the material and instructors. Live chat functionality significantly enhanced the learning experience, underlining its importance in the user interface. Students asked questions on the topics covered, asked for help, and even started discussions on the future of AI with the instructors.

Students' progress over the learning material was primarily tracked through (a) the completion rate and (b) grades on quizzes, the former being used as a proxy to measure engagement. While we found that a mark as completed button helped gamify the learning experience, it should be noted that the completion rate is an imperfect metric as it does not necessarily yield a grasp of the material. In order to improve the reliability of this metric, we aim to incorporate additional checks in the platform in the future, such as only allowing students to mark a section as completed if they viewed all of the material and answered all of the quiz questions.

Covering foundational modules first and foremost was instrumental in preparing students for more complex topics. Furthermore, we discovered that delving into the history of AI captivated students, with a particular interest in early AI milestones like Shakey the first robot and logical agents, providing valuable context for understanding modern deep learning systems.

However, we encountered challenges finding a suitable coding interface, with Google Colab ultimately not being an optimal medium for programming instruction for this age group. Students faced technical barriers using Google Colab, running code cells, installing libraries, and troubleshooting errors. We found this medium to be lacking the necessary interactivity for this age group and it was challenging for students to collaborate with their peers. It was also difficult for instructors to provide instantaneous feedback and hands-on guidance.

Addressing prerequisite knowledge gaps, particularly in probability, logic, and mathematics, emerged as a crucial aspect of effective instruction. Moreover, integrating ethical considerations into the curriculum raised students' awareness of AI's impact on fairness, safety, vulnerability, and explainability. The accessibility of mobile applications was highlighted, as some students used mobile phones and tablets to access the platform.

Overall, this experience showcased the potential for innovative education methods and online learning platforms for teaching AI to teenagers.

\section{Conclusion and Future Work}
\label{conclusion}

The titanic task of bringing college-level AI instruction to teenagers is challenging. It requires an immense mobilization from different parties to make the AI curriculum understandable and accessible to younger learners.  
While considering ways to bridge this gap, two central tracks emerged. The formal education track will incorporate AI in schools. Understandably, this will take some time to realize. The second track is to adopt an informal AI education route, leveraging any opportunities outside the classroom to achieve this goal. This may be the most feasible and imminent way to spread AI knowledge among teens.

Based on the survey results, completion rates, engagement, and
quiz grades, we can deem our study a successful experience for our team to engage a cohort of high school students with foundational AI concepts. It gave us confidence that informal learning is not only a feasible but also an effective approach. One key finding is that teenagers are more than capable of grasping AI concepts despite the complexities inherent to the subject.

In pursuing our mission, our future informal AI interventions will consider the wealth of lessons we learned from our experience. First, we will design larger-scale AI bootcamps to collect solid empirical evidence for further research and assessment of the learning outcomes of these bootcamps. Encouraging teamwork and fostering creativity is another aspect we want to emphasize in future efforts. 

Overall, we acknowledge the importance of pursuing rigorous research for informal learning to identify pathways for learning AI, adapting to the varying levels of computing proficiency and different interest levels among students.  


\bibliography{aiphabet}

\newpage

\section{Appendix}

\end{document}